\documentclass[aps,pre,twocolumn,showpacs,showkeys,groupedaddress,amsmath,twoside]{revtex4}
\usepackage{graphicx,dsfont}

\defcitealias{krapivsky1996life}{[P.~L.~Krapivsky and S.~Redner, \emph{Am.~J.~Phys.~}\textbf{64}, 546 (1996)]}
\defcitealias{turban1992anisotropic}{[L.~Turban, \emph{J.~Phys.~A~}\textbf{25}, 127 (1992)]}

\newcommand{\e}{\mathrm{e}}
\newcommand\FF[3]{%
  {\vphantom{#2}}#1#2#3%
}

\begin{document}

%\preprint{}

\title{Weighted {K}olmogorov-{S}mirnov test:\\ Accounting for the tails}


\author{R\'emy~Chicheportiche$^{1,2}$ and Jean-Philippe~Bouchaud$^1$}
%\thanks{}
\affiliation{
$^1$ Capital~Fund~Management, 75\,007 Paris, France\\
$^2$ Chaire de finance quantitative, Ecole Centrale Paris, 92\,295 Ch\^atenay-Malabry, France
}


\date{October 8, 2012}%\today}

\begin{abstract}
Accurate goodness-of-fit tests for the extreme tails of empirical distributions is a very important issue, 
relevant in many contexts, including geophysics, insurance, and finance. 
We have derived exact asymptotic results for a generalization of the large-sample Kolmogorov-Smirnov test, 
well suited to testing these extreme tails. 
In passing, we have rederived and made more precise the approximate limit solutions found originally in unrelated fields,
first in \citetalias{turban1992anisotropic} and later in \citetalias{krapivsky1996life}.
%first in [L.~Turban, \emph{J.~Phys.~A~}\textbf{25}, 127 (1992)] and later in [P.~L.~Krapivsky and S.~Redner, \emph{Am.~J.~Phys.~}\textbf{64}, 546 (1996)].
\end{abstract}

\pacs{02.50.Ey, 03.65.Ge, 05.10.Gg}
\keywords{Goodness-of-Fit tests, Kolmogorov-Smirnov, Survival probability, Particle in a box}

\maketitle

%%%%%%%%%%%%%%%%%%%%%%%%%%%%%%%%%%%%%%%%%%%%%%%%%%%%%%%%%%%%%%%%%%%%%%%%
\section{Introduction and motivation}

The problem of testing whether a null-hypothesis theoretical probability distribution
is compatible with the empirical probability distribution of a sample of observations
is known as goodness-of-fit (GoF) testing and is ubiquitous in all fields of science and engineering.
The best known theoretical result is due to Kolmogorov and Smirnov (KS) \cite{kolmogorov1933sulla,smirnov1948table}, 
and has led to the eponymous statistical test.  
Several specific cases have been studied (and/or are still under scrutiny), including: 
univariate or multivariate samples \cite{fasano1987multidimensional,cabana1994goodness,cabana1997transformed,Fermanian2005119}, 
independent or dependent data \cite{chicheportiche2011goodness}, different choices of distance measures \cite{darling1957kolmogorov},
investigation of different parts of the distribution domain \cite{anderson1952asymptotic,deheuvels2009weighted}, etc. 

This class of problems has a particular appeal for physicists since the works of Doob \cite{doob1949heuristic} and Khmaladze \cite{khmaladze1982martingale},
who showed how GoF testing is related to stochastic processes. Finding the law of a test often amounts to treating a Fokker-Planck problem,
which in turn maps into a Schr\"odinger equation for a particle in a certain potential confined by walls. 

The classical KS test suffers from an important flaw: the test is only weakly sensitive to the quality of the fit in the tails of the tested
distribution, when it is often these tail events (corresponding to centennial floods, devastating earthquakes, financial crashes, etc.) 
that one is most concerned with (see, e.g., Ref.~\cite{clauset2009power}). 
Here we focus on a GoF test for a univariate sample of size $N\gg 1$, with the Kolmogorov distance but equi-weighted quantiles, 
which is equally sensitive to all regions of the distribution. 
We unify two earlier attempts at finding asymptotic solutions, one by Anderson and Darling in 1952 \cite{anderson1952asymptotic} 
and a more recent, seemingly unrelated one that deals with ``life and death of a particle in an expanding cage'' 
by Krapivsky and Redner \cite{krapivsky1996life,redner2007guide}. 
We present here the exact asymptotic solution of the corresponding stochastic problem, 
and deduce from it the precise formulation of the GoF test, which is of a fundamentally different nature than the KS test.

\section{Empirical cumulative distribution and its fluctuations}

Let $\mathbf{X}$ be a latent random vector of $N$ independent and identically distributed variables, 
with marginal cumulative distribution function (cdf) $F$. 
One realization of $\mathbf{X}$ consists of a time series $\{x_1,\ldots,x_n, \ldots, x_N\}$ that exhibits no persistence (see 
Ref.~\cite{chicheportiche2011goodness} when some non trivial dependence is present).
For a given number $x$ in the support of $F$, let $\mathbf{Y}(x)$ be the random vector 
the components of which are the Bernoulli variables ${Y}_n(x)=\mathds{1}_{\{X_n\leq x\}}$.
%The expected value and the covariance of $\mathbf{Y}_n(x)$ are given by:
The one-point and two-points expectations of ${Y}_n(x)$ are
\begin{eqnarray*}
	\mathds{E}[{Y}_n(x)]&=&F(x),\\
    \mathds{E}[{Y}_n(x){Y}_m(x')]&=&\begin{cases}F(\min(x,x'))&,\,n=m\\F(x)F(x')&,\,n\neq m\end{cases}.
\end{eqnarray*}
The centered sample mean of $\mathbf{Y}(x)$ is: 
\begin{equation}\label{eq:Ybar}
	\overline{{Y}}(x)=\frac{1}{N}\sum_{n=1}^N{Y}_n(x)-F(x),
\end{equation}
which measures the difference between the empirically determined cdf at point $x$ and its true value.
It is therefore the quantity on which any statistics for GoF testing is built.
Denoting $u=F(x)$ and $v=F(x')$, the covariance function of $\overline{{Y}}$ is easily shown to be:
\[
	{\rm Cov}(\overline{{Y}}(u),\overline{{Y}}(v))=\frac{1}{N}\big(\min(u,v)-uv\big),
\]
where now and in the following 
\begin{equation}\tag{\ref{eq:Ybar}$'$}
	\overline{{Y}}(u)=\frac{1}{N}\sum_{n=1}^N{Y}_n(F^{-1}(u))-u.
\end{equation}

\subsection*{Limit properties}
One now defines the process ${y}(u)$ as the limit of $\sqrt{N}\,\overline{{Y}}(u)$ when $N \to \infty$.
For a given $u$, it represents the difference between the empirically determined cdf 
of the (infinitely many) $X$'s and the theoretical one, evaluated at the $u$-th quantile.
According to the Central Limit Theorem, it is Gaussian and its covariance function is given by:
\begin{equation}\label{eq:Htheo}
	{I}(u,v)=\min(u,v)-uv,
\end{equation}
which characterizes the so-called Brownian bridge, i.e.\  a Brownian motion ${y}(u)$  such that \mbox{$y(u\!=\!0)=y(u\!=\!1)=0$}. 

Interestingly, $F$ does not appear in Eq.~(\ref{eq:Htheo}) anymore, 
so the law of any functional of the limit process $y$ is independent of the law of the underlying finite size sample.
This property is important for the design of \emph{universal} GoF tests.

\subsection*{Norms over processes\\and the {K}olmogorov-{S}mirnov test}
In order to measure a limit distance between distributions, a norm $||.||$ over the space of continuous bridges needs to be chosen. 
Typical such norms are the norm-2 (or `Cramer-von Mises' distance)
\[
	||{y}||_2=\int_0^1y(u)^2du,
\]
as the bridge is always integrable, or the norm-sup 
\[
	||{y}||_\infty=\sup_{u\in[0,1]}|y(u)|,
\]
as the bridge always reaches an extremal value (also called the Kolmogorov distance).
Unfortunately, both these norms mechanically overweight the core values $u \approx 1/2$ and disfavor the tails $u \approx 0,1$:
since the variance of $y(u)$ is zero at both extremes and maximal in the central value, the major contribution to $||{y}||$ indeed
comes from the central region. To alleviate this effect, in particular when the GoF test is intended to investigate 
a specific region of the domain, it is preferable to introduce additional weights and study $||y\sqrt{\psi}||$ rather than $||y||$ itself.
Anderson and Darling show in Ref.~\cite{anderson1952asymptotic} that the solution to the problem with the Cramer-von Mises norm and arbitrary weights $\psi$
is obtained by spectral decomposition of the covariance kernel, and use of Mercer's theorem. 
In this note we will rather focus on the case of the weights $\psi$ being equal to the inverse variance
$\psi(u)=1/\mathds{V}[y(u)]$, which equi-weights all quantiles, and with the Kolmogorov distance. 

Solutions for the distributions of such variance-weighted Kolmogorov-Smirnov statistics were studied by No\'e,
leading to the laws of the one-sided \cite{noe1968calculation} and two-sided \cite{noe1972calculation} finite sample tests.
They were later generalized and tabulated numerically by H.~Niederhausen \cite{niederhausen1981sheffer,niederhausen1981tables,wilcox1989percentage}.
However, although exact and appropriate for small samples, these solutions rely on recursive relations and are not in closed form.
We instead come up with an analytic closed-form solution for large samples that relies on an elegant analogy from statistical physics.

\section{The weighted {B}rownian bridge: Law of the supremum}

So again $y(u)$ is a Brownian bridge, i.e.\ a centered Gaussian process on $u\in[0,1]$ with covariance function
${I}(u,v)$ given in Eq.~(\ref{eq:Htheo}).
In particular, $y(0)=y(1)=0$ with probability equal to 1, no matter how distant $F$ is from the sample cdf around the core values.
In order to zoom on these tiny differences in the tails, we weight the Brownian bridge as follows: 
for given $a\in]0,1[$ and $b\in [a,1[$, we define
\begin{equation}\label{eq:weightedY}
	\tilde{y}(u)\equiv y(u)\sqrt{\psi(u;a,b)},
\end{equation}
with
\[
	\psi(u;a,b)=\left\{\begin{array}{ll}\frac{1}{u(1-u)}&,\,a\leq u\leq b\\0&,\,\text{otherwise.}\end{array}\right.
\]
We will characterize the law of the supremum $K(a,b)\equiv\sup_{u\in[a,b]}\left|\tilde{y}(u)\right|$:
\begin{align*}\nonumber
	\mathcal{P}_{\!{\scriptscriptstyle <}}(k|a,b)   &\equiv\mathds{P}[K(a,b)\leq k]\\
                                                    &=\mathds{P}\left[|\tilde{y}(u)|\leq k, \forall u\in[a,b]\right].
\end{align*}

\subsection*{Diffusion in a cage with moving walls}

Define the time change $t=\frac{u}{1-u}$. The variable $W(t)=(1+t)\,y\!\left(\frac{t}{1+t}\right)$  
is then a Brownian motion (Wiener process) on $[\tfrac{a}{1-a},\tfrac{b}{1-b}]$, since one can check that:
\[
	{\rm Cov}\big(W(t),W(t')\big)=\min(t,t').
\]
$\mathcal{P}_{\!{\scriptscriptstyle <}}(k|a,b)$ can be now written as
\[
	\mathcal{P}_{\!{\scriptscriptstyle <}}(k|a,b)=\mathds{P}\left[|W(t)|\leq k\sqrt{t},\forall t\in[\tfrac{a}{1-a},\tfrac{b}{1-b}]\right].
\]

The problem with initial time $\frac{a}{1-a}=0$ and horizon time $\frac{b}{1-b}=T$ 
has been treated by Krapivsky and Redner in Ref.~\cite{krapivsky1996life}
as the survival probability $S(T;k=\sqrt{\frac{A}{2D}})$ of a Brownian particle 
diffusing with constant $D$ in a cage with walls expanding as $\sqrt{At}$. 
Their result is that for large $T$, 
\[
S(T;k)\equiv\mathcal{P}_{\!{\scriptscriptstyle <}}(k|0,\tfrac{T}{1+T})\propto T^{-\theta(k)}.
\]
They obtain analytical expressions for $\theta(k)$ in both limits \mbox{$k\to 0$} and \mbox{$k\to\infty$}.
The limit solutions of the very same differential problem were found earlier by Turban for the critical behavior of the directed self-avoiding walk in parabolic 
geometries \cite{turban1992anisotropic}.

We take here a slightly different route, suggested by Anderson and Darling in Ref.~\cite{anderson1952asymptotic} 
but where the authors did not come to a conclusion. Our contributions are: 
(i) we treat the general case $a>0$ for \emph{any} $k$;
(ii) we explicitly compute the $k$-dependence of both the exponent \emph{and} the prefactor of the power-law decay; and
(iii) we provide the link with the theory of GoF tests and 
compute the pre-asymptotic distribution when $]a,b[\to]0,1[$ of the weighted Kolmogorov-Smirnov test statistics.

Choosing a constant weight function $\psi$ instead of the one above corresponds to the usual KS case and leads, 
after appropriate change of variable and time change, to a similar problem of a Brownian diffusion 
inside a box with walls moving at \emph{constant} velocity. Since the walls now expand as $Vt$ faster than the 
diffusive particle can move, the survival probability clearly decays to a positive value. 
The resulting survival probability turns out to be the usual Kolmogorov-Smirnov distribution.
Other choices of $\psi$ generally result in much harder problems. % see Ref.~\cite{anderson1952asymptotic}.

Still, a simple and elegant GoF test for the tails \emph{only} can be designed starting with digital weights in the form
$\psi(u;a)=\mathds{1}_{\{u\geq a\}}$ or $\psi(u;b)=\mathds{1}_{\{u\leq b\}}$ for upper and lower tail, respectively.
The corresponding test laws can be read off Eq.~(5.9) in Ref.~\cite{anderson1952asymptotic}.
\footnote{The quantity $M$ appearing there is the volume under the normal bivariate surface between specific bounds,
and it takes a very convenient form in the unilateral cases $\tfrac{1}{2}\leq a \leq u \leq 1$ and $0\leq u \leq b\leq\frac{1}{2}$.
Mind the missing $j$ exponentiating the alternating $(-1)$ factor.}
Investigation of both tails is attained with
$\psi(u;q)=\mathds{1}_{\{u\leq 1-q\}}+\mathds{1}_{\{u\geq q\}}$ (where $q>\tfrac{1}{2}$).

\subsection*{An {O}rnstein-{U}hlenbeck process with fixed walls} 

Introducing now the new time change $\tau=\ln\sqrt{\frac{1-a}{a}\,t}$, the variable $Z(\tau)=W(t)/\sqrt{t}$ is a stationary 
Ornstein-Uhlenbeck process on $[0,T]$ 
where 
\begin{equation}\label{eq:Tchange}
    T=\ln\sqrt{\frac{b(1-a)}{a(1-b)}},
\end{equation}
and
\[
	{\rm Cov}\big(Z(\tau),Z(\tau')\big)=\e^{-|\tau-\tau'|}.
\]
Its dynamics is described by the stochastic differential equation
\begin{equation}
	dZ(T)=-Z(T)dT+\sqrt{2}\,dB(T),
\end{equation}
with $B(T)$ an independent Wiener process.
The initial condition for $T=0$ (corresponding to $b=a$) is $Z(0)=y(a)/\sqrt{\mathds{V}[y(a)]}$,
a random Gaussian variable of zero mean and unit variance. The distribution $\mathcal{P}_{\!{\scriptscriptstyle <}}(k|a,b)$ can now be understood as 
the unconditional survival probability of a mean-reverting particle in a cage with fixed absorbing walls:
\begin{align*}
	%S(T;k) =
    \mathcal{P}_{\!{\scriptscriptstyle <}}(k|T)
	      &=\mathds{P}\left[-k\leq Z(\tau)\leq k, \forall \tau\in[0,T]\right]\\
	      &=\int_{-k}^{k}f_T(z;k)dz,
\end{align*}
where 
\[
    f_T(z;k)dz=\mathds{P}\big[Z(T)\in[z,z+dz[ \mid \{Z(\tau)\}_{\tau<T}\big]
\]
is the density probability of the particle being at $z$
at time $T$, when walls are in $\pm k$. Its dependence on $k$, although not explicit on the right hand side, is due to the boundary condition 
associated with the absorbing walls (it will be dropped in the following for the sake of readability) \footnote{
In particular, $\mathcal{P}_{\!{\scriptscriptstyle <}}(k|0)=\operatorname{erf}\left(\tfrac{k}{\sqrt{2}}\right)$.}.

The Fokker-Planck equation governing the evolution of the density $f_T(z)$ reads
\[
	\partial_{\tau}f_{\tau}(z)=\partial_z\left[z\,f_{\tau}(z)\right]+\partial_z^2\left[f_{\tau}(z)\right],\quad 0<\tau\leq T.
\]
Calling $\mathcal{H}_{\text{FP}}$ the second order differential operator $-\left[\mathds{1}+z\partial_z+\partial_z^2\right]$, 
the full problem thus amounts to finding the general solution of
\[
    \bigg\{
	\begin{array}{rcl}
	-\partial_{\tau}f_\tau(z)&=&\mathcal{H}_{\text{FP}}(z)f_\tau(z)\\
	f_{\tau}(\pm k)&=&0, \forall \tau\in[0,T]
	\end{array}
    \bigg. .
\]
We have explicitly introduced a minus sign since we expect that the density decays with time in an absorption problem.
Because of the term $z\partial_z$, $\mathcal{H}_{\text{FP}}$ is not hermitian and thus cannot be diagonalized.
However, as is well known, one can define $f_{\tau}(z)=\e^{-\frac{z^2}{4}}\phi_{\tau}(z)$ and the Fokker-Planck equation becomes
\[
    \bigg\{
	\begin{array}{rcl}
	-\partial_{\tau}\phi_\tau(z)&=&\left[-\partial_z^2+\frac{1}{4}z^2-\frac{1}{2}\mathds{1}\right]\phi_{\tau}(z)\\
	\phi_{\tau}(\pm k)&=&0, \forall \tau\in[0,T]
	\end{array}
    \bigg. ,
\]
and its Green's function, i.e.\ the (separable) solution \emph{conditionally on the initial position} $(z_{\text{i}},T_{\text{i}})$, 
is the superposition of all modes
\[
	G_{\phi}(z,T\mid z_{\text{i}},T_{\text{i}})=\sum_{\nu}\e^{-\theta_{\nu}(T-T_{\text{i}})}\widehat\varphi_{\nu}(z)\widehat\varphi_{\nu}(z_{\text{i}}),
\]
where $\widehat\varphi_{\nu}$ are the normalized solutions of the stationary Schr\"odinger equation
\[
    \Bigg\{
	\begin{array}{rcl}
	\left[-\partial_z^2+\frac{1}{4}z^2\right]\varphi_{\nu}(z)&=&\left(\theta_{\nu}+\frac{1}{2}\right)\varphi_{\nu}(z)\\
	\varphi_{\nu}(\pm k)&=&0
	\end{array}
    \Bigg.,
\]
each decaying with its own energy $\theta_\nu$, 
where $\nu$ labels the different solutions with increasing eigenvalues,
and the set of eigenfunctions $\{\widehat\varphi_{\nu}\}$ defines an orthonormal basis of the Hilbert space on which $\mathcal{H}_{\text{S}}(z)=\left[-\partial_z^2+\frac{1}{4}z^2\right]$ acts.
In particular, 
\begin{equation}\label{eq:orthonormality}
	\sum_{\nu}\widehat\varphi_{\nu}(z)\widehat\varphi_{\nu}(z')=\delta(z-z'),
\end{equation}
so that indeed $G(z,T_{\text{i}}\mid z_{\text{i}},T_{\text{i}})=\delta(z-z_{\text{i}})$, and the general solution writes
\begin{align*}
	f_T(z_T;k)%&=\int_{-k}^{k}\frac{\e^{-\frac{z_T^2}{4}}}{\e^{-\frac{z_{\text{i}}^2}{4}}}G_{\phi}(z_T,T\mid z_{\text{i}},T_{\text{i}})\,f_0(z_{\text{i}})dz_{\text{i}}\\\nonumber
		       &=\int_{-k}^{k}\e^{\frac{z_{\text{i}}^2-z_T^2}{4}}G_{\phi}(z_T,T\mid z_{\text{i}},T_{\text{i}})\,f_0(z_{\text{i}})dz_{\text{i}},
\end{align*}
where $T_{\text{i}}=0$, which corresponds to the case $b=a$ in Eq.~(\ref{eq:weightedY}), and $f_0$ is the distribution of the initial value $z_{\text{i}}$
which is here, as noted above, Gaussian with unit variance.

$\mathcal{H}_{\text{S}}$ figures out an harmonic oscillator of mass $\frac{1}{2}$ and frequency $\omega=\frac{1}{\sqrt{2}}$ within 
an infinitely deep well of width $2k$: its eigenfunctions are parabolic cylinder functions \cite{mei1983harmonic,gradshteyn1980table}
\begin{align*}
    y_+(\theta;z)&=\phantom{z\,}\e^{-\frac{z^2}{4}}\,\FF{_1}{F}{_1}\!\left(-\tfrac{  \theta}{2},\tfrac{1}{2},\tfrac{z^2}{2}\right)\\
    y_-(\theta;z)&=         z\, \e^{-\frac{z^2}{4}}\,\FF{_1}{F}{_1}\!\left( \tfrac{1-\theta}{2},\tfrac{3}{2},\tfrac{z^2}{2}\right)
\end{align*}
properly normalized. 
The only acceptable solutions for a given problem are the linear combinations of $y_+$ and $y_-$ 
which satisfy orthonormality (\ref{eq:orthonormality})  and the boundary conditions:
for periodic boundary conditions, only the integer values of $\theta$ would be allowed, 
whereas with our Dirichlet boundaries \mbox{$|\widehat\varphi_{\nu}(k)|=-|\widehat\varphi_{\nu}(-k)|=0$}, real non-integer eigenvalues $\theta$ are allowed.
\footnote{A similar problem with a \emph{one-sided} barrier leads to a continuous spectrum; 
this case has been studied originally in Ref.~\cite{mei1983harmonic} and more recently in Ref.~\cite{lladser2000domain} 
(it is shown that there exists a quasi-stationary distribution for any $\theta$)
and generalized in Ref.~\cite{aalen2004survival}.
} 
For instance, the fundamental level $\nu=0$ is expected to be the symmetric solution 
\mbox{$ \widehat\varphi_0(z)\propto y_+(\theta_0;z)$}
with $\theta_0$ the smallest possible %non-odd 
value compatible with the boundary condition:
\begin{equation}
	\theta_0(k)=\inf_{\theta>0}\big\{\theta:y_+(\theta;k)=0\big\}.
\end{equation}
In what follows, it will be more convenient to make the $k$-dependence explicit, and
a hat will denote the solution with the normalization relevant to our problem, 
namely 
%$\widehat{\varphi}_{\nu}(z;k)=\varphi_{\nu}(z;k)/||\varphi_{\nu}||_k$,
 $\widehat{\varphi}_{0  }(z;k)=y_+(\theta_0(k);z)/||y_+||_k$,
with the norm
\[
   %||\varphi_{\nu}||_k^2 \equiv\int_{-k}^k\varphi_{\nu}(z;k)^2dz
    ||y_+||_k^2 \equiv\int_{-k}^ky_+(\theta_0(k);z)^2dz,
\]
so that
$%\[
    \int_{-k}^k\widehat{\varphi}_{\nu}(z;k)^2dz=1.
$%\]

\subsection*{Asymptotic survival rate}

Denoting by $\Delta_{\nu}(k)\equiv[\theta_{\nu}(k)-\theta_0(k)]$ the gap between the excited levels and the fundamental, 
the higher energy modes $\widehat\varphi_\nu$ cease to contribute to the Green's function when $\Delta_{\nu}T\gg 1$, 
and their contributions to the above sum die out exponentially as $T$ grows.
Eventually, only the lowest energy mode $\theta_0(k)$ remains, and the solution tends to 
\[
	f_{T}(z;k)=A(k)\,\e^{-\frac{z^2}{4}}\,\widehat\varphi_{0}(z;k)\,\e^{-\theta_{0}(k)T},
\]
when $T\gg (\Delta_1)^{-1}$, with 
\begin{equation}\label{eq:A_tilde}
	A(k)=\int_{-k}^k\e^{\frac{z_{\text{i}}^2}{4}}\widehat\varphi_0(z_{\text{i}};k)f_0(z_{\text{i}})dz_{\text{i}}.
\end{equation}

Let us come back to the initial problem of the weighted Brownian bridge reaching its extremal value in $[a,b]$.
If we are interested in the limit case where $a$ is arbitrarily close to $0$ and $b$ close to $1$, then $T \to \infty$ and
the solution is thus given by
\begin{align*}
	%S(T;k)  
    \mathcal{P}_{\!{\scriptscriptstyle <}}(k|T)&=A(k)\,\e^{-\theta_{0}(k)T}\int_{-k}^{k}\e^{-\frac{z^2}{4}}\widehat\varphi_{0}(z;k)dz\\
            &=\widetilde{A}(k)\, \e^{-\theta_{0}(k)T},  
\end{align*}
with $\widetilde{A}(k) \equiv \sqrt{2\pi}A(k)^2$.

We now compute explicitly the limit behavior of both $\theta_0(k)$ and $\widetilde{A}(k)$.
\paragraph*{$\boxed{k\to \infty}$} 
As $k$ goes to infinity, the absorption rate $\theta_0(k)$ is expected to converge toward $0$: intuitively, an infinitely 
far barrier will not absorb anything. At the same time, $\mathcal{P}_{\!{\scriptscriptstyle <}}(k|T)$ must tend to 1 in that limit. 
So $\widetilde{A}(k)$ necessarily tends to unity. Indeed,
\begin{align}
	       \theta_0(k)&\xrightarrow{k\to\infty} \sqrt{\frac{2}{\pi}}k\,\e^{-\frac{k^2}{2}} \to 0,\\\nonumber
	  \widetilde{A}(k)&\xrightarrow{k\to\infty}\left(\int_{-\infty}^{\infty}\widehat{\varphi}_{0}(z;\infty)^2dz\right)^2 = 1.
\end{align}
In principle, we see from Eq.~(\ref{eq:A_tilde}) that corrections to the latter arise both (and jointly) from 
the functional relative difference of the solution 
\mbox{$\epsilon(z;k)=y_+(\theta_0(k);z)/y_+(0;z)-1$}, 
and from the finite integration limits ($\pm k$ instead of $\pm \infty$).
However, it turns out that the correction of the first kind is of second order in $\epsilon$. 
\footnote{
From Eq.~(\ref{eq:A_tilde}) we have, when \mbox{$k\to\infty$},
\[
    A(k)=(2\pi)^{-1/2}\frac{\int_{-k}^k\e^{-z^2/2}[1+\epsilon(z;k)]dz}{\sqrt{\int_{-k}^k\e^{-z^2/2}[1+\epsilon(z;k)]^2dz}}.
\]
The result follows by keeping only the dominant terms in the expansion in powers of $\epsilon(z;k)$.
A similar computation for the asymptotic analysis by expanding the wave function in $\theta$ was performed in Ref.~\cite{krapivsky1996life}.
Alternatively, algebraic arguments allow to understand that, to first order in the energy correction $\theta_0(k)-\theta_0(\infty)$, 
the perturbation of the wave function is orthogonal to $\widehat{\varphi}_{0}(z;\infty)$.} 
The correction to $A(k)$ 
is thus dominated by the finite integration limits $\pm k$, so that% pre-asymptotically:
\begin{equation}
    \widetilde{A}(k\to\infty)\approx \operatorname{erf}\left(\frac{k}{\sqrt{2}}\right)^2.
\end{equation}
\paragraph*{$\boxed{k\to 0}$}
For small $k$, the system behaves like a free particle in a sharp and infinitely deep well, since the quadratic potential is almost flat around 0.
The fundamental mode becomes then 
\[
	\widehat\varphi_0(z;k\to 0)=\frac{1}{\sqrt{k}}\cos \left(\frac{\pi z}{2k}\right),
\]
and consequently
\begin{align}
	       \theta_0(k)&\xrightarrow{k\to 0}\frac{\pi^2}{4k^2}-\frac{1}{2},\\\nonumber
	  \widetilde{A}(k)&\xrightarrow{k\to 0}\left(\int_{-k}^{k}\frac{\e^{-\frac{z^2}{4}}}{(2\pi)^{\frac{1}{4}}}
	  \frac{1}{\sqrt{k}}\cos\left(\frac{\pi z}{2k}\right)dz\right)^2\\
                        &\approx\frac{1}{\sqrt{2\pi}k}
	  \left(\frac{4k}{\pi}\right)^2=\frac{16}{\pi^2\sqrt{2\pi}}k.
\end{align}

\begin{figure}[!h!]
	\center
	\includegraphics[scale=0.8,trim=0 0 -25 0,clip]{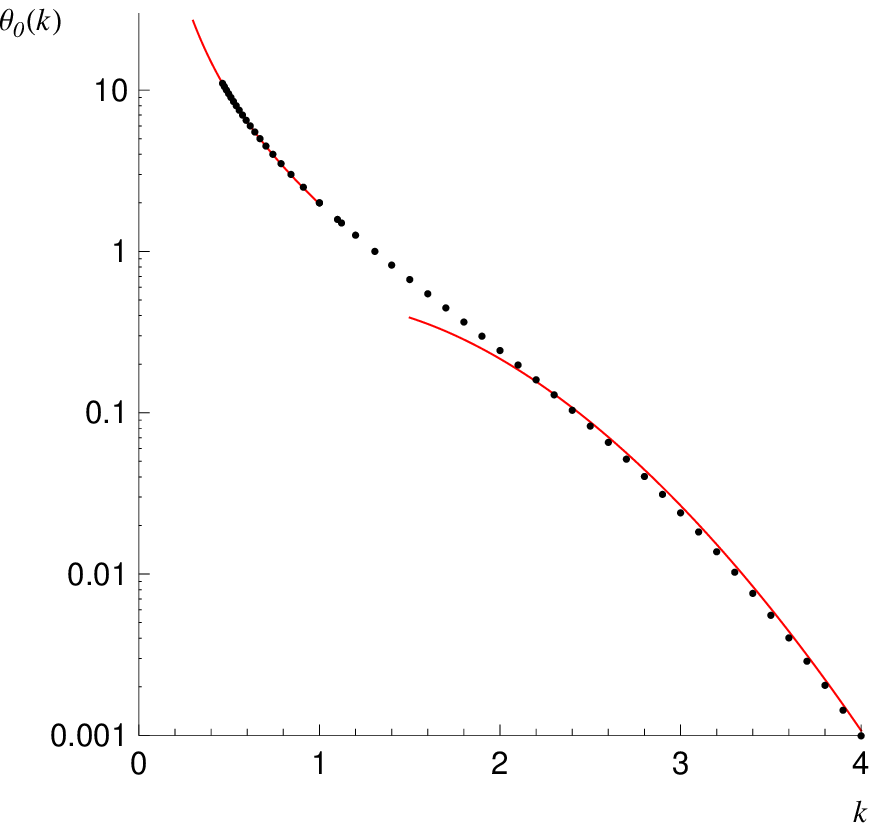}
	\includegraphics[scale=0.8,trim=0 0 -25 0,clip]{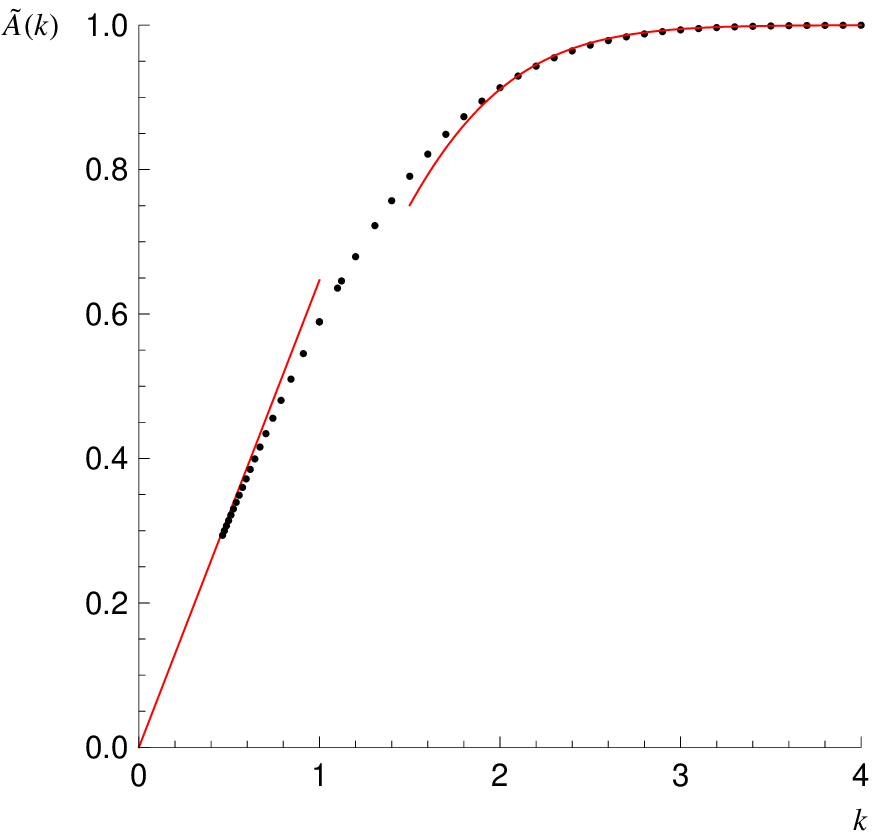}
	\caption{\textbf{Top:    } Dependence of the exponent       $\theta_0$ on $k$; similar to Fig.~2 in Ref.~\cite{krapivsky1996life}, but in lin-log scale; see in particular Eqs.~(9b) and (12) there.
	         \textbf{Bottom: } Dependence of the prefactor $\widetilde{A}$ on $k$. The red solid lines illustrate the analytical behavior in the
	         limiting cases $k\to 0$ and $k\to\infty$.}
	\label{fig:A_theta}
\end{figure}

We show in Fig.~\ref{fig:A_theta} the functions $\theta_0(k)$ and $\widetilde{A}(k)$ computed numerically from the exact solution,
together with their asymptotic analytic expressions. In intermediate values of $k$ (roughly between 0.5 and 3) these limit
expressions fail to reproduce the exact solution.

\subsection*{Higher modes and validity of\\ the asymptotic ($N\gg 1$) solution}
\begin{figure}[!h!]
	\center
	\includegraphics[scale=0.85,trim=0 0 -25 0,clip]{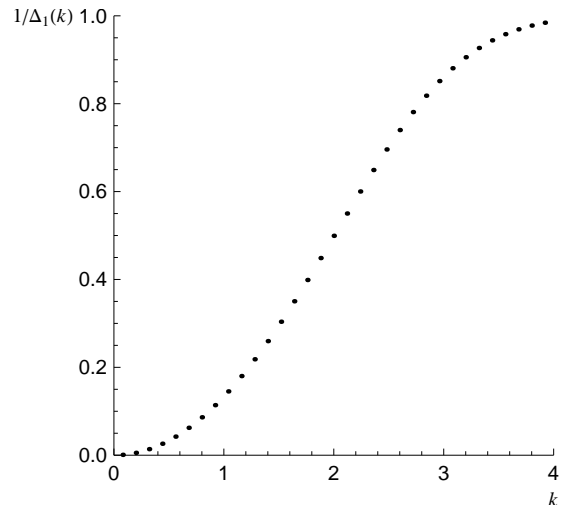}
	\caption{$1/\Delta_1(k)$ saturates to 1, so that the condition \mbox{$N\gg\exp[1/\Delta_1(k)]$} is virtually always satisfied.}
	\label{fig:delta1}
\end{figure}

Higher modes $\nu>0$ with energy gaps $\Delta_\nu\lesssim 1/T$ must in principle be kept in the pre-asymptotic computation.
This, however, is irrelevant in practice since the gap \mbox{$\theta_1-\theta_0$} is never small.
Indeed, $\widehat\varphi_1(z;k)$ is proportional to the asymmetric solution $y_-(\theta_1(k);z)$
and its energy
\[
    \theta_1(k)=\inf_{\theta>\theta_0(k)}\big\{\theta:y_-(\theta;k)=0\big\}
\]
is found numerically to be very close to \mbox{$1+4\theta_0(k)$}. 
In particular, $\Delta_1>1$ (as we illustrate in Fig.~\ref{fig:delta1}) 
and thus $T \Delta_1 \gg 1$ will always be satisfied in cases of interest. 

\section{Back to {GoF} testing and conclusion}

Let us now come back to GoF testing. In the case of a constant weight, corresponding to the classical KS test, the probability 
\mbox{$\mathcal{P}_{\!{\scriptscriptstyle <}}(k|a\!=\!0,b\!=\!1)$} is well defined and has the well known KS form \cite{kolmogorov1933sulla}:
\[
\mathcal{P}_{\!{\scriptscriptstyle <}}(k|a=0,b=1) = 1 - 2 \sum_{n=1}^{\infty} (-1)^{n-1} \e^{-2 n^2 k^2},
\]
which, as expected, grows from $0$ to $1$ as $k$ increases. The value $k^*$ such that this probability is $95 \%$ is $k^* \approx 1.358$ 
\cite{smirnov1948table}. This can be interpreted as follows: if, for a data set of size $N$, the maximum value of $\overline{{Y}}(u)$ is larger
than $\approx 1.358/\sqrt{N}$, then the hypothesis that the proposed distribution is a ``good fit'' can be rejected with $95 \%$ confidence.

In order to convert the above calculations into a meaningful test, one must specify values of $a$ and $b$. The natural choice is $a=1/N$ and 
$b=1-a$, corresponding to the min and max of the sample series. Indeed, 
$a=F(\min z)\approx \frac{1}{N}\sum_{n=1}^N\mathds{1}_{\{z_n\leq\min z\}}=\frac{1}{N}$, and similarly for $b$. Correspondingly, the 
relevant value of $T$ is given, according to Eq.~(\ref{eq:Tchange}) above, by 
\[
	T=\ln\sqrt{\frac{b(1-a)}{a(1-b)}} \approx \ln N,\qquad N \gg 1.
\]
This leads to our central result for the cdf of the weighted maximal Kolmogorov distance $K(\tfrac{1}{N\!+\!1},\tfrac{N}{N\!+\!1})$ 
under the hypothesis that the tested and the true distributions coincide:
\begin{equation}\label{eq:final_res}
	\boxed{S(N;k)=\mathcal{P}_{\!{\scriptscriptstyle <}}(k|\ln N)=\widetilde{A}(k)N^{-\theta_0(k)}},
\end{equation}
which is valid whenever $N \gg 1$ since, as we discussed above, the energy gap $\Delta_1$ is greater than unity.

\begin{figure}
	\center
	\includegraphics[scale=0.85,trim=0 10 -25 0,clip]{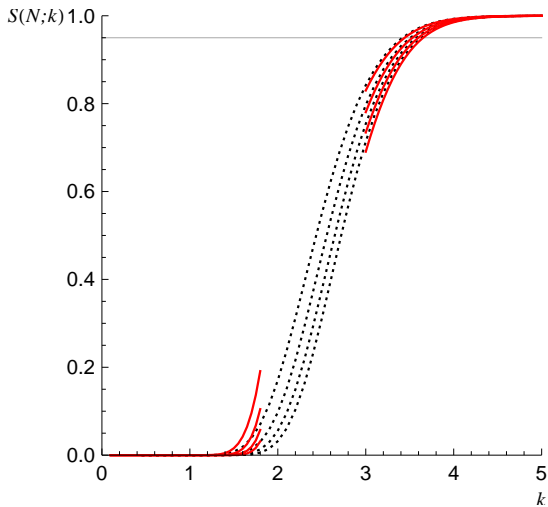}
	\caption{Dependence of $S(N;k)$ on $k$ for $N=10^3,10^4,10^5,10^6$ (from left to right).
	         As $N$ grows toward infinity, the curve is shifted to the right, and eventually $S(\infty;k)$ is zero for any $k$.
	         The red solid lines illustrate the analytical behavior in the
	         limiting cases $k\to 0$ and $k\to\infty$. The horizontal grey line corresponds to a $95 \%$ confidence level.}
	\label{fig:S}
\end{figure}

The final cumulative distribution function (the test law) is depicted in Fig.~\ref{fig:S} for different values of the sample size $N$.
Contrarily to the standard KS case, this distribution \emph{still depends on $N$}. 
In particular, the threshold value $k^*$ corresponding to a $95 \%$ confidence level increases with $N$. 
Since for large $N$, $k^* \gg 1$ one can use the asymptotic expansion above, which soon becomes quite accurate, as shown in Fig.~\ref{fig:S}. 
This leads to:
\[
\theta_0(k^*) \approx - \frac{\ln 0.95}{\ln N} \approx \sqrt{\frac{2}{\pi}}k^*\,\e^{-\frac{k^{*2}}{2}},
\]
which gives $k^* \approx 3.439, 3.529, 3.597, 3.651$ for, respectively, $N=10^3,10^4,10^5,10^6$. 
For exponentially large $N$ and to logarithmic accuracy, one has: $k^* \sim \sqrt{2 \ln (\ln N)}$. 
This variation is very slow, but one sees that as a matter of principle, the ``acceptable'' maximal 
value of the weighted distance is much larger (for large $N$) than in the KS case.

%%% CODE R 
%%% fon <- function(k,N) {sqrt(2/pi)*k*exp(-k^2/2)+log(.95)/log(N)}
%%% round(sapply(10^(3:6),function(N) uniroot(fon,c(3,4),N=N)$root),3)

In conclusion, we believe that accurate GoF tests for the extreme tails of empirical distributions is a very important issue, relevant in many 
contexts. We have derived exact asymptotic results for a generalization of the Kolmogorov-Smirnov test, 
well suited to testing the whole domain up to these extreme tails. 
Our final results are summarized in Eq.~(\ref{eq:final_res}) and Fig.~\ref{fig:S}. 
In passing, we have rederived and made more precise the result of 
Krapivsky and Redner \cite{krapivsky1996life} concerning the survival probability of a diffusive particle in an expanding cage. It would be 
interesting to exhibit other choices of weight functions that lead to soluble survival probabilities. It would also be interesting to extend the 
present results to multivariate distributions, and to dependent observations, along the lines of Ref.~\cite{chicheportiche2011goodness}.

We want to thank Sid Redner for a useful discussion and for his inspiring work, and Lo\"ic Turban for bringing Ref.~\cite{turban1992anisotropic} to our attention.

\bibliography{biblio_all}

\end{document}